\documentstyle[aps,epsf,twocolumn,amsfonts]{revtex}

\def\Jb{\bar{J}}

\def\be{\begin{equation}}
\def\ee{\end{equation}}
\def\bdm{\begin{displaymath}}
\def\edm{\end{displaymath}}
\def\bea{\begin{eqnarray}}
\def\eea{\end{eqnarray}}
\def\nn{\nonumber\\}

\def\r#1{(\ref{#1})}

\def\psib{\skew5\bar\psi}

\def\sb{{\bar s}}
\def\s{\sigma}

\def\bra#1{\langle #1 |}
\def\ket#1{|#1\rangle}

\begin{document}
\draft
\title{The fate of spinons in spontaneously dimerised
spin-${1\over 2}$ ladders}
\author{Dave Allen$^1$, Fabian H.L. Essler$^1$ and Alexander A. Nersesyan$^{2,3}$}
\address{
$^1$Department of Physics, Theoretical Physics, Oxford 
University\\
1 Keble Road, Oxford, OX1 3NP, United Kingdom\\
$^2$ICTP, P.O. Box 586, 34100 Trieste, Italy\\
$^3$The Andronikashvili Institute of Physics, 
Tamarashvili
6, Tbilisi, Georgia.}
\date{\today}
\maketitle
\begin{abstract}
We study a weakly coupled, frustrated two-leg spin-1/2 Heisenberg
ladder. For vanishing coupling between the chains, elementary
excitations are deconfined, gapless spin-1/2 objects called {\sl
spinons}. We investigate the fate of spinons for the case of a weak
interchain interaction. 
We show that despite a drastic change in ground state, which becomes
spontaneously dimerised, spinons survive as elementary excitations but
acquire a spectral gap. We furthermore determine the {\sl exact}
dynamical structure factor for several values of momentum transfer.
\end{abstract}
\narrowtext
\section{Introduction}
The role of frustration in quasi one-dimensional magnetic materials
has attracted much experimental and theoretical attention in recent
years. 
On the theoretical side, the simplest example of a frustrated quantum
magnet is the spin-1/2 Heisenberg antiferromagnetic chain with nearest
neighbour exchange $\delta J$ and next-nearest neighbour exchange
$J$. This model is equivalent to a two-leg ladder (see
Fig.1), where the coupling along (between) the legs of
the ladder is equal to $J$ ($\delta J$). 

\begin{figure}[ht]
\begin{center}
\noindent
\epsfxsize=0.45\textwidth
\epsfbox{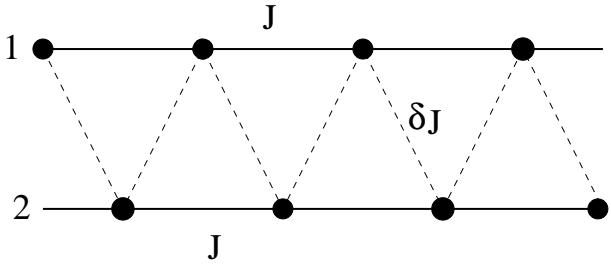}
\end{center}
\caption{\label{fig:ladder}%
Heisenberg zigzag ladder.
} 
\end{figure}

The zigzag ladder model is believed to describe the quantum magnet
${\rm SrCuO_2}$ \cite{srcuo2,matsuda} above the magnetic ordering
transition, which takes place at about $T\approx 2K$. The exchange
constants are estimated to be $J\approx 1800K$, $|\delta J/J|\approx
0.1-0.2$ \cite{matsuda}.
A second material with zigzag structure, that has recently
attracted much interest, is ${\rm Cs_2CuCl_4}$
\cite{radu}. However, in ${\rm Cs_2CuCl_4}$ all neighbouring chains
are coupled by a zigzag interaction and no pronounced ladder
structure exists.

In Refs. \cite{hal,wa,allen} it was argued that a weak
antiferromagnetic zigzag coupling between the chains drives the model
to a massive phase, characterised by spontaneous dimerisation (see
also \cite{ft}). Let us briefly review some important parts of the
derivations of \cite{wa,allen}. The lattice Hamiltonian of the zigzag
ladder is
\bea
{ H}&=& J\sum_{j=1,2}\sum_{n}
{\bf S}_{j,n}\cdot {\bf S}_{j,n+1}\nn
&&+ \delta J\sum_{n}\left(
{\bf S}_{1,n}+
{\bf S}_{1,n+1}\right)
\cdot
{\bf S}_{2,n}\ ,
\label{hamilspin}
\eea
where we assume that $\delta\ll 1$. The low-energy effective action
for \r{hamilspin} is now obtained as follows. For $\delta\to 0$ one is
dealing with two decoupled Heisenberg chains, which can be bosonised
in terms of two Wess-Zumino-Novikov-Witten (WZNW) models by using the
standard relation between the spin density on chain $j$ and the fields of
the WZNW model (see e.g. \cite{affleck2})
\bea
{S^a_j(x)\over a_0}&=&\left[ J^a_j(x)+\Jb^a_j(x)\right]
+(-1)^{x/a_0} n^a_j(x)\ .
\label{spinWZW}
\eea
Here $a_0$ is the lattice spacing, and the fields $J^a_j$ and $\Jb^a_j$ are
the right and left currents of the WZWN model corresponding to chain
$j$. They parametrise the smooth component of the
magnetisation. Finally, ${\vec n}_j$ is the staggered component of the
magnetisation on chain $j$. Using (\ref{spinWZW}), the zigzag interchain
interaction can be expressed in terms of the WZNW fields. 
In this way one straightforwardly obtains the current-current
interaction \cite{wa,allen}
\bea
{\cal H}_{c}=
\lambda_1(J_1^a+\Jb_1^a)(J_2^a+\Jb_2^a)\nn
-\lambda_0(J_1^a\Jb_1^a+J_2^a\Jb_2^a)\ ,
\label{hcc}
\eea
where $\lambda_1\propto \delta J$. A standard renormalisation group
(RG) analysis then shows that the antiferromagnetic interchain
interaction $\lambda_1$ leads to a spontaneously dimerised ground
state \cite{wa,allen}.
In \cite{NGE} it was shown that, in addition to the current-current
interaction \r{hcc}, a ``twist'' term arises
\be
{\cal H}_{t}=\rho (n^a_1\partial_x n^a_2
-n^a_2\partial_x n^a_1)\ .
\label{twist}
\ee
In the presence of exchange anisotropies the twist term induces
incommensurabilities in the spin correlations \cite{NGE}. We expect
this to be hold true even in the SU(2) symmetric case (no exchange
anisotropies) we are interested in here.
In the latter case it can be shown that the twist term and current-current
interaction are equally important in the RG sense: they diverge
(i.e. reach strong coupling) simultaneously, with a fixed ratio
\cite{unpub}. As far as the SU(2) symmetric zigzag ladder is
concerned, it is therefore not possible to separate the effects of the
twist and current-current interactions in a simple way.

However, from a purely theoretical point of view it clearly is desirable
to develop a thorough understanding of the physics due to isolated
current-current and twist interactions. Their effects can be
disentangled by introducing an exchange anisotropy \cite{NGE},
which makes the twist more and the current-current interaction less
relevant in the RG sense. Using this trick, a pure twist interaction
was studied in \cite{NGE}.

The role of an isolated current-current interaction has been
previously investigated in connection with the zigzag ladder in
\cite{ft,wa,allen}. In particular, the spectrum of elementary
excitations and the dynamical structure factor were calculated in
\cite{allen} using large-N techniques. It is known, that extrapolation
of large-N results to small values of N can lead to incorrect results
\cite{karow}. Having this in mind, we carry out an {\sl exact}
calculation in order to determine the spectrum and structure factor.
We find that the large-N results are indeed qualitatively incorrect.

As explained above, in the zigzag ladder both twist and
current-current interactions are present; therefore,
strictly speaking, our results cannot be directly
applied to this model. Nevertheless, we believe that many of our findings 
presented below remain
qualitatively correct when applied to model (\ref{hamilspin}). 
We discus this point in more detail in section
\ref{section:summary}. 

In order to connect our results to a microscopic model, we 
consider a frustrated spin ladder modified in such a way that only
current-current interactions emerge in the low-energy effective action. 

The outline of this paper is as follows: in section \ref{section:model}
we introduce a frustrated spin ladder model giving rise to the desired
low-energy effective field theory. In section \ref{section:duality} we
show that the resulting field theory is essentially equivalent to an O(4)
Gross-Neveu model \cite{GN}. 
Sections \ref{section:gs} and \ref{section:exc} are
concerned with the description of the ground state(s) and elementary 
excitations. In section \ref{section:structure} we determine
the exact dynamical structure factor for several values of momentum
transfer and show that there are no coherent contributions to the
structure factor. We conclude with a summary and discussion of our
results. 

\section{A frustrated ladder without twist}
\label{section:model}

The model we consider is a generalisation \cite{WEI} of the standard
two-leg spin ladder which, apart from the on-rung coupling $J_{\perp}$,
also includes an interaction $J_{\times}$ across both diagonals of the
plaquettes. The Hamiltonian reads
\bea
H &=& J \sum_{j=1,2} \sum_n {\bf S}_{j,n} \cdot {\bf S}_{j,n + 1}
+ J_{\perp} \sum_n {\bf S}_{1,n} \cdot {\bf S}_{2,n} \nonumber\\
&+& J_{\times} \sum_n \left[  {\bf S}_{1,n} \cdot {\bf S}_{2,n + 1}
+ {\bf S}_{1,n + 1} \cdot {\bf S}_{2,n} \right]
\label{ham}
\eea
We assume that
\be
J, J_{\perp}, J_{\times} > 0, ~~~~
J \gg J_{\perp}, J_{\times}\ .
\ee
\begin{figure}[ht]
\begin{center}
\noindent
\epsfxsize=0.45\textwidth
\epsfbox{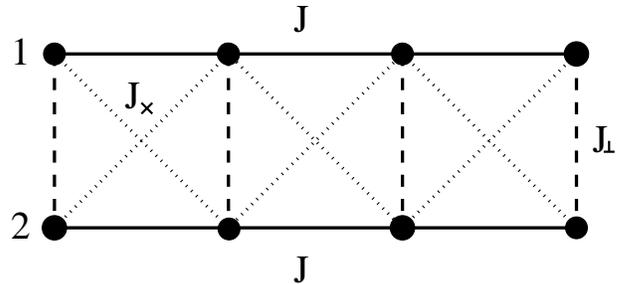}
\end{center}
\caption{\label{fig:ladder2}%
The twistless ladder model.
} 
\end{figure}
The low-energy effective action can be derived by nonabelian
bosonisation in the usual way. The Hamiltonian density is found to be
of the form
\be
{\cal H} (x) = {\cal H}_1 (x) + {\cal H}_2 (x) + {\cal H}_{\em int}
(x)\ ,
\label{ham-dens}
\ee
where ${\cal H}_{1,2}$  are critical $SU_1(2)$ WZNW models with a
marginally irrelevant current-current perturbation ($\lambda_0 > 0$):
\bea
{\cal H}_j &=& \frac{2\pi v_s}{3} \left( :{\bf \Jb}_{j} \cdot {\bf \Jb}_{j}: + 
:{\bf J}_{j} \cdot {\bf J}_{j}: \right)\nn
&& - \lambda_0 {\bf \Jb}_{j} \cdot {\bf J}_{j}\ ,\quad j = 1,2.
\label{WZW}
\eea
The interaction part is given by
\bea
{\cal H}_{\em int}  &=& \lambda_1 \left({\bf J}_1+{\bf\Jb}_1\right)
 \cdot  \left({\bf J}_2+{\bf\Jb}_2\right)
+\lambda_2 {\bf n}_1 \cdot {\bf n}_2\ ,
\label{Hint}
\eea
with the coupling constants 
\be
\lambda_1 = (J_{\perp} + 2 J_{\times}) a_0 , ~~~
\lambda_2 = (J_{\perp} - 2 J_{\times}) a_0.
\ee
No marginal perturbation with the twist-term structure 
arises because the staggered magnetisation operators add
rather than subtract due to the geometry of the problem. The absence
of such term can also be deduced from the existence of discrete
(reflection) symmetries of the lattice Hamiltonian \r{ham}.
If $J_\perp= 2J_\times$ only the marginal (current-current) interaction
survives. This is the case we study in the remainder of this paper.

We note that for generic values of $J_\perp$ and $J_\times$ the
interaction of staggered magnetisations dominates and the resulting
physics is essentially the same as for the standard ladder
($J_\times=0$) \cite{SNT} (see also chapter 21 of \cite{GNT}).

\section{Duality Transformation}
\label{section:duality}

The low-energy effective action \r{WZW}-\r{Hint} can be recast as a
theory of four massive, interacting, real (Majorana) fermions, or
equivalently, four weakly coupled Ising models \cite{allen} 
\bea
&&{\cal H}=\frac{i}{2}\sum_{\alpha=0}^3{v_\alpha}
(\psi_\alpha\partial_x\psi_\alpha-\psib_\alpha\partial_x\psib_\alpha)\nn
&&+\frac{\lambda_1-\lambda_0}{2}\sum_{j>i=1}^3
\psi_i\psib_i\psi_j\psib_j
-\frac{\lambda_1+\lambda_0}{2}\psi_0\psib_0\sum_{i=1}^3\psi_i\psib_i.
\label{anisot}
\eea
Here $v_1=v_2=v_3=v_s\neq v_0$ are the velocities of the four Majorana
fermions.
The lattice spin operators are expressed in terms of the Majorana
fields and order and disorder operators of the four Ising models as
\bea
S^z_+(x)&\propto&
-i\left(\psi_1\psi_2+\psib_1\psib_2\right)
- {\cal A} (-1)^{x/a_0}\mu_1\mu_2\s_3\s_0\ ,\nn
S^z_-(x)&\propto& i\left(\psi_3\psi_0+\psib_3\psib_0\right)
+ {\cal A} (-1)^{x/a_0}\s_1\s_2\mu_3\mu_0\ ,
\eea
where $S^z_\pm(x)=S^z_1(x)\pm S^z_2(x)$  and ${\cal A}$ is a
nonuniversal constant. Analogous expressions are available for the
other components of the spin operators \cite{allen}.
A standard one-loop RG analysis shows that the
coupling $\lambda_0$ flows to zero, so we will ignore it in what
follows. In order to further simplify the problem, we also
neglect the small difference between the velocities $v_s$ and $v_0$,
and finally perform a duality transformation on the 0-Majorana 
\be
\psi_0\rightarrow \psi_4\ , \psib_0\rightarrow -\psib_4\ ,
\s_0\rightarrow \mu_4\ , \mu_0\rightarrow \s_4\ .
\label{dual.tr}
\ee
This yields the Hamiltonian of the $O(4)$ Gross-Neveu model \cite{GN}
\begin{equation}
{\cal H}=\frac{iv_s}{2}\sum_{i=1}^4\psi_i\partial_x\psi_i
-\psib_i\partial_x\psib_i+\frac{\lambda_1}{2}\sum_{j>i=1}^4
\psi_i\psib_i\psi_j\psib_j\ .
\label{grossneveu}
\end{equation}
Under (\ref{dual.tr})
the spin-densities transform to
\bea
S^z_+(x)&\propto&
-i\left(\psi_1\psi_2+\psib_1\psib_2\right)
- {\cal A} (-1)^{x/a_0}\mu_1\mu_2\s_3\mu_4\ ,\nn
S^z_-(x)&\propto& i\left(\psi_3\psi_4-\psib_3\psib_4\right)
+ {\cal A} (-1)^{x/a_0}\s_1\s_2\mu_3\s_4\ .
\eea
\section{Ground State}
\label{section:gs}
In order to proceed, it is convenient to use the representation 
of \r{grossneveu} in terms of two sine-Gordon models \cite{ws}.
Ignoring terms that only renormalise the velocity we find that
(\ref{grossneveu}) is equivalent to
\bea
&&{\cal H}=\sum_{i=\pm}{v_s\over 2}\left[(\partial_x\varphi_i)^2
+(\partial_x\theta_i)^2\right]\nn
&&+2\lambda_1\left[\frac{1}{8\pi}
\left((\partial_x\varphi_i)^2-(\partial_x\theta_i)^2\right)
-\frac{1}{(2\pi a_0)^2}\cos\sqrt{8\pi}\varphi_i\right],\nn
\label{SG}
\eea
where $\theta_i$ are the dual fields. The two sine-Gordon models
\r{SG} occur on the SU(2) invariant strong-coupling separatrix of the
Kosterlitz-Thouless phase diagram and are thus in the massive regime.

\subsection{Twistless Ladder}

The low-energy effective model \r{SG} exhibits a
local $Z_2$ symmetry related to {\sl independent} translations
by one lattice spacing on each chain ($\varphi_\pm\rightarrow
\varphi_\pm+\sqrt{\pi/2}$). This symmetry is spontaneously broken
in the ground state and leads to a nonvanishing dimerization.
Notice that the $Z_2$ symmetry appears to be a feature 
of the low-energy sector only and follows from the fact that spin
currents $\vec{J}_{1,2}$ are translationally invariant objects.
The transformation ${\bf S}_1 (n)\rightarrow \frac{1}{2}
\left[{\bf S}_{1,n+1}+{\bf S}_{1,n-1}\right]$, or a similar one
with ${\bf S}_{1,k} \rightarrow {\bf S}_{2,k}$,
changes the lattice Hamiltonian but leaves the low-energy 
effective field theory invariant and maps the two ground 
states onto one another. 

In order to characterise the dimerisation
patterns of the two ground states, we determine the expectation values
\bea
\langle \vec{S}_{1,n}\cdot \vec{S}_{2,n}\rangle&\propto&
\langle \vec{J}_{1}(x)\cdot \vec{J}_{2}(x)\rangle
+\langle \vec{n}_{1}(x)\cdot \vec{n}_{2}(x)\rangle\ ,\nn
\langle \vec{S}_{1,n}\cdot \vec{S}_{2,n+1}\rangle&\propto&
\langle \vec{J}_{1}(x)\cdot \vec{J}_{2}(x)\rangle
-\langle \vec{n}_{1}(x)\cdot \vec{n}_{2}(x)\rangle\ ,\nn
\langle \vec{S}_{2,n}\cdot \vec{S}_{1,n+1}\rangle&\propto&
\langle \vec{J}_{1}(x)\cdot \vec{J}_{2}(x)\rangle
-\langle \vec{n}_{1}(x)\cdot \vec{n}_{2}(x)\rangle\ .
\label{SS}
\eea
After performing the duality transformation to the O(4) Gross-Neveu
model and bosonising, we obtain
\bea
\langle \vec{n}_{1}(x)\cdot
\vec{n}_{2}(x)\rangle&\propto&
\langle\cos\sqrt{2\pi}\varphi_{+}
\cos\sqrt{2\pi}\varphi_{-}\rangle =\pm{\rm const}\  m\ ,\nn
\langle \vec{J}_{1}(x)\cdot
\vec{J}_{2}(x)\rangle&\propto&
\langle\left(\cos\sqrt{2\pi}\varphi_{+}
\cos\sqrt{2\pi}\varphi_{-}\right)^2\rangle\nn
&&\qquad ={\rm const}\  m^2\ ,
\label{dimer}
\eea
where $m\propto\exp(-{\rm const}J/J_\perp)$ is 
the (exponentially small) soliton mass in the Sine-Gordon model.
Due to the smallness of $m$, 
the $\langle \vec{n}_{1}(x)\cdot\vec{n}_{2}(x)\rangle$ 
expectation value dominates in (\ref{SS}), so that
within the exponential accuracy
the dimerisation is proportional to the quantum soliton mass. 


\begin{figure}[ht]
\begin{center}
\noindent
\epsfxsize=0.45\textwidth
\epsfbox{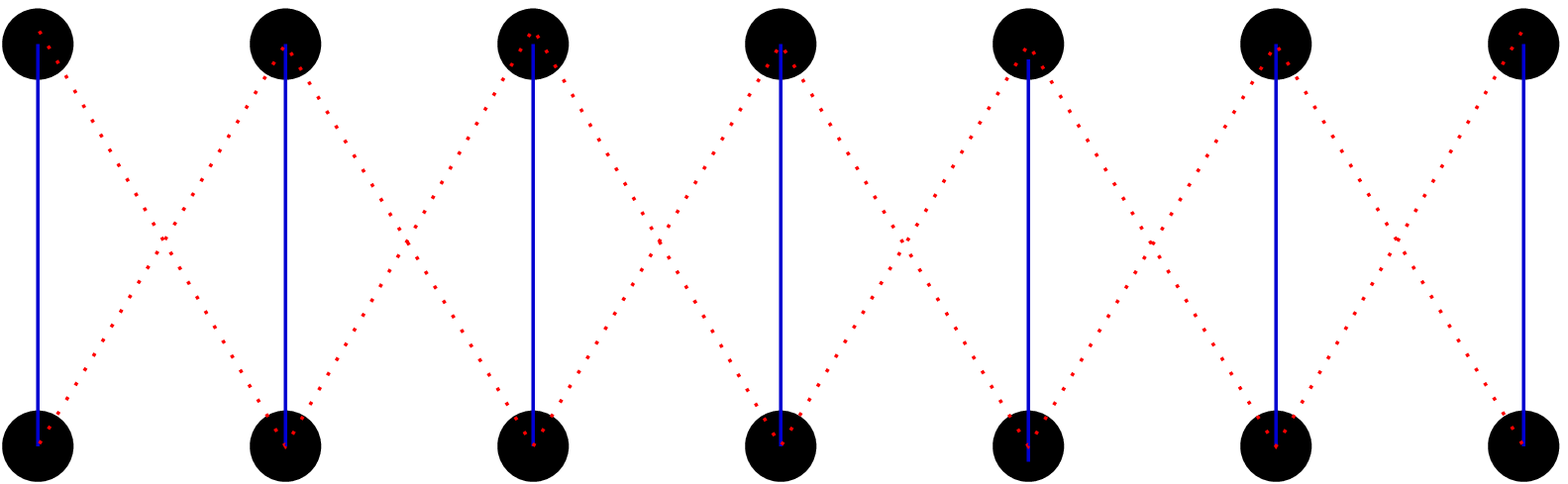}
\end{center}
\vskip .5cm
\begin{center}
\noindent
\epsfxsize=0.45\textwidth
\epsfbox{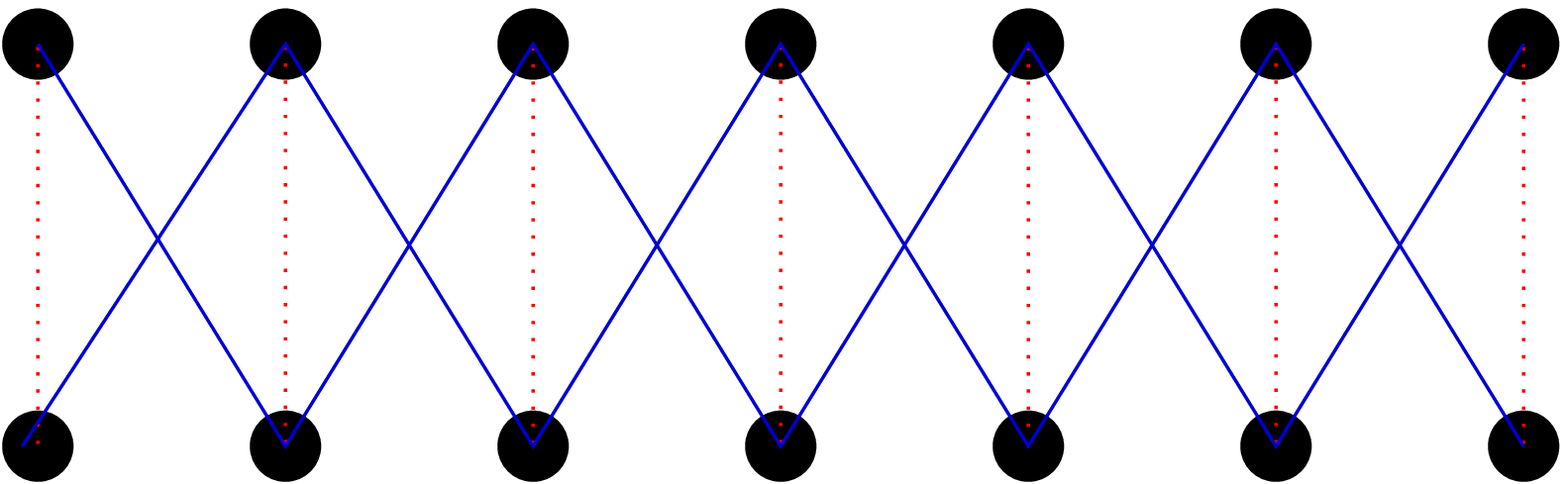}
\end{center}
\caption{\label{fig:groundstate}%
Qualitative picture of the two degenerate dimerised ground states: 
spins connected by the solid (dotted) lines have a tendency to form
singlets (triplets).
} 
\end{figure}
The $Z_2$ symmetry of the low-energy effective Hamiltonian \r{SG},
that manifests itself in the degeneracy of the two ground states
corresponding to different signs in \r{dimer}, is spontaneously
broken, implying the existence of massive $Z_2$ kinks. It turns out
(see below) that these kinks are elementary excitations of the model.
\subsection{Zig-Zag Ladder}

\begin{figure}[ht]
\begin{center}
\noindent
\epsfxsize=0.4\textwidth
\epsfbox{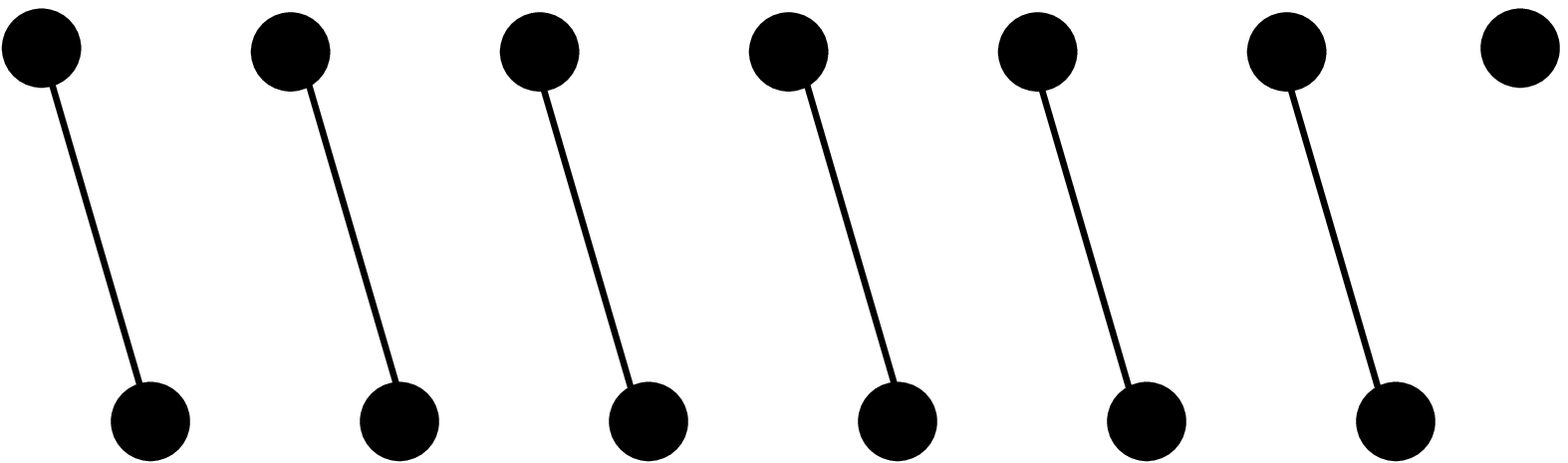}
\end{center}
\vskip .5cm
\begin{center}
\noindent
\epsfxsize=0.4\textwidth
\epsfbox{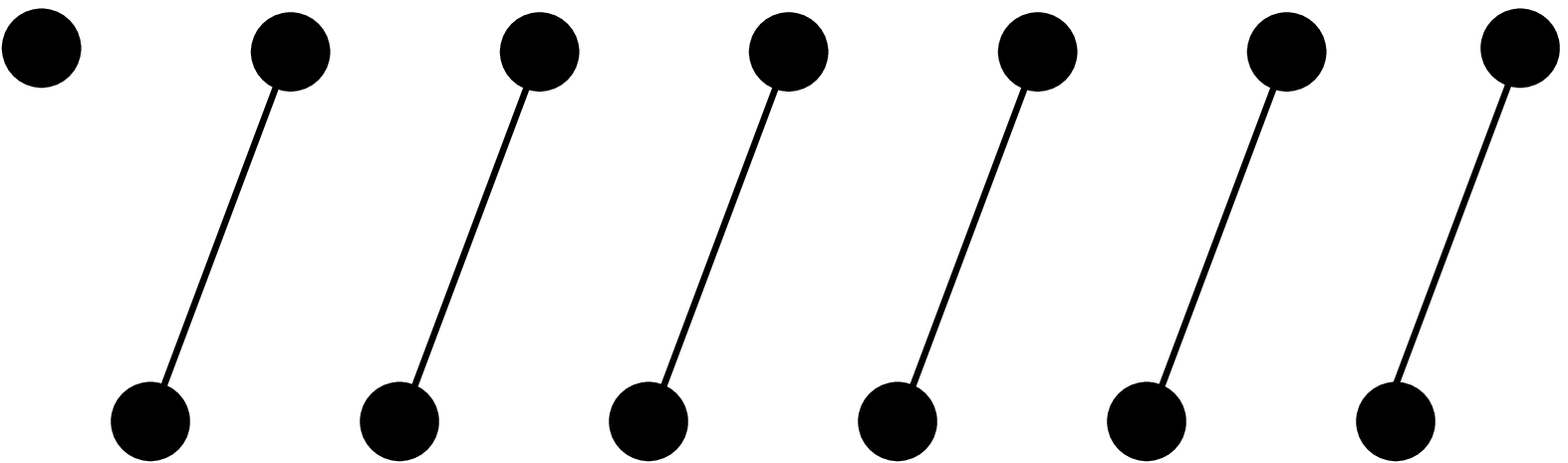}
\end{center}
\caption{\label{fig:class1}%
Qualitative picture of the spin configuration in the dimerised ground
states: spins along the thick diagonal bonds have a tendency to form
singlets. 
} 
\end{figure}

Let us discuss the implications of the emergence of spontaneous
dimerisation for the case of the zig-zag ladder if we ignore the twist
term. For the zig-zag ladder the appropriate definition for the
dimerisation
is
\be
d=\left\langle {\vec S}_1(x)\cdot\left({\vec S}_2(x+a_0/2)
-{\vec S}_2(x-a_0/2)\right)\right\rangle .
\ee
In the continuum limit we find
\bea
d&\propto& \langle\cos\sqrt{2\pi}\varphi_{+}
\cos\sqrt{2\pi}\varphi_{-}\rangle =\pm{\rm const}\  m\ .
\label{dimer2}
\eea
The resulting dimerisation patterns are shown in Fig.\ref{fig:class1}.
We believe that taking into account the twist term will not
qualitatively change this picture.

\section{Excitations}
\label{section:exc}

{From} the exact solution of the Sine-Gordon models \r{SG} we infer
that there are only four elementary excitations corresponding to
solitons and antisolitons in the $\pm$ sectors. We denote these by
$s_\pm$ and $\bar{s}_\pm$. The elementary excitation have a simple
interpretation in terms of {\sl dimerisation kinks}, i.e. domain walls
separating regions of dimerisation with opposite sign.
It can be shown along the lines of \cite{topch} that these particles 
carry spin $\pm 1/2$. In terms of the low-energy effective theory
of two Sine-Gordon models \r{SG}, the total spin density is given by
\be
S^z_1(x)+S^z_2(x)=\frac{1}{\sqrt{2\pi}}\left[
\partial_x\varphi_+(x)+\partial_x\varphi_-(x)\right].
\label{totspin}
\ee
Kinks interpolate between asymptotic values of the fields 
$\varphi_i$ differing by $\pm\sqrt{\pi/2}$ as is most easily
deduced from the fact that the classical vacua of \r{SG} are located at
\be
\langle \varphi_i\rangle_{\rm class}=
\sqrt{\frac{\pi}{2}}n_i\ ,\quad i=\pm\ ,
\ee
where $n_i$ are arbitrary integers. Integration of \r{totspin} then yields 
that a single kink carries spin
\be
S^z=\pm\frac{1}{\sqrt{2\pi}}\sqrt{\frac{\pi}{2}}=\pm\frac{1}{2}.
\ee
The results presented below for the dynamical
structure factor are consistent with the interpretation of these kinks
as {\sl gapped spinons}. Altogether there are two spin-$1/2$
multiplets, corresponding to one multiplet for each leg of the ladder. 
The emerging physical picture is quite simple and pretty: the two-spinon
states observed in the structure factor simply correspond to the kinks
related to the spontaneous breakdown of the discrete $Z_2$ symmetry. 

\begin{figure}[ht]
\begin{center}
\noindent
\epsfxsize=0.4\textwidth
\epsfbox{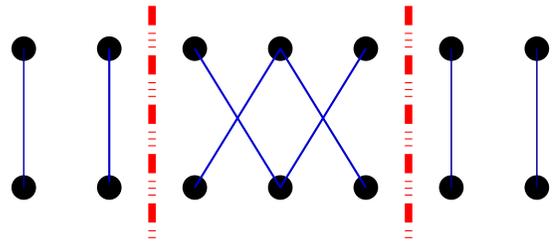}
\end{center}
\caption{\label{fig:exc}%
A two-spinon state in the twistless ladder. Spinons correspond to kinks
between domains with different sign of the dimerisation. Solid
lines depict bonds along which there is a tendency to form singlets.
} 
\end{figure}

Simple visualizations of this picture are shown in Fig.\ref{fig:exc} for
the twistless ladder and in Fig.\ref{fig:class2} for the zig-zag ladder.

\begin{figure}[ht]
\begin{center}
\noindent
\epsfxsize=0.4\textwidth
\epsfbox{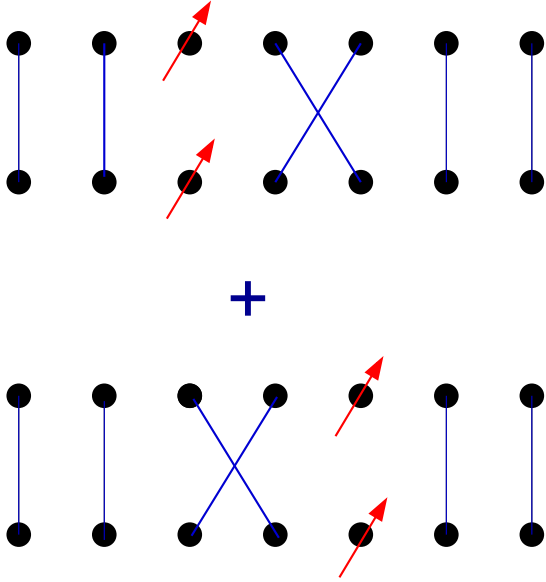}
\end{center}
\caption{\label{fig:exc2}%
``Resonating'' ideal dimer configurations.
} 
\end{figure}

\begin{figure}[ht]
\begin{center}
\noindent
\epsfxsize=0.4\textwidth
\epsfbox{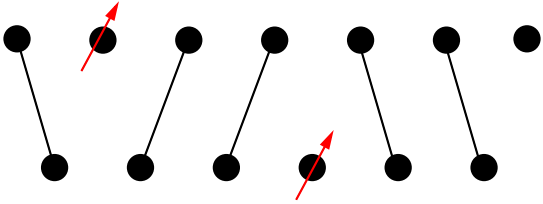}
\end{center}
\caption{\label{fig:class2}%
Physical picture of a two-spinon state. Spinons correspond to kinks
connecting domains with different sign of the dimerisation.
} 
\end{figure}

For the twistless ladder the kinks correspond to vertical domain walls
between regions with different signs of dimerisation. There is a
spin-1/2 associated with each domain wall, although this is not
immediately obvious from Fig.\ref{fig:exc}. In order to get a feeling
why a spin-1/2 might be associated with each kink, let us think of the
translationally invariant, ``double-zigzag'' ground state shown in
Fig.\ref{fig:groundstate} as a symmetric superposition of two
dimerised states. Each such state represents a sequence of plaquettes
with ideal singlet bonds across the plaquette diagonals
(with each spin involved in one bond only), has a period
$2a_0$ and is shifted with respect to the other state by one lattice
spacing. If the ``double-zigzag'' phase occupies a finite
domain of the ladder, for the two $2a_0$-periodic dimerised states to
resonate, the number of rungs within such a domain should be odd.
Then the two-kink configuration in Fig.\ref{fig:exc} can equivalently
be viewed as the superposition of states shown in Fig.\ref{fig:exc2}.

The intuitive picture one obtains from Fig.\ref{fig:exc2} is then that
on average there is indeed a spin-1/2 associated with each kink.

For the zig-zag case a much nicer picture emerges. The $Z_2$ symmetry
corresponds to a reflection symmetry on the lattice and kinks look
like left over spin-1/2's as shown in Fig.\ref{fig:class2}.

The intuitive picture of Fig.~\ref{fig:class2} fits well to the
identification of a spinon in a spin-1/2 chain as a bare spin insertion
into the ground state \cite{talstra}. 

\section{Dynamical structure factor}
\label{section:structure}
The long-distance asymptotics of the spin-spin correlation functions are
dominated by the soft modes at $q=0,\pi$, $q_\perp=0,\pi$, where $q$
and $q_\perp$ denote the wave-numbers along and perpendicular to the
two chains, respectively. In what follows we will determine the
dynamical structure factor for wave numbers in the vicinity of the
above four points in $\vec{q}$-space.
Due to the spin-rotational symmetry the dynamical structure factor is
given by 
\bea
&&S(\omega,q,q_\perp)\propto{\rm Im}i\int_{-\infty}^\infty
dx\int_0^\infty dt
e^{-i\omega t+iqx}\nn
&&\times\langle\left[\{S_1^z(t,x)\pm S_2^z(t,x)\}
,\{S^z_1(0,0)\pm S^z_2(0,0)\}\right]\rangle\ .
\eea
where the positive (negative) sign corresponds to $q_\perp=0$
($q_\perp=\pi$). 

\subsection{Summary of large-N results}
The dynamical structure factor has been previously calculated in the
framework of a large-$N$ approach \cite{allen}. 
The limit $N \rightarrow \infty$ of (\ref{grossneveu}) is equivalent
to a theory of free massive Majorana fermions 
\begin{equation}
{\cal H}=\frac{iv_s}{2}\sum_{i=1}^N\psi_i\partial_x\psi_i
-\psib_i\partial_x\psib_i+im\psi_i\psib_i\ .
\label{largeN}
\end{equation}
The presence of the mass term reflects the spontaneous breakdown of
parity, which in turn implies the existence of two degenerate ground 
states. The sign of the mass terms, as well as 
the expectation values $\langle \s_a\rangle$ and $\langle\mu_a\rangle$, 
depend on the choice of ground state. In the case where
$\langle\s_a \rangle\neq 0$ (the mass of the triplet is positive),
the structure factor for $q_\perp=0$ and $q$ around $0,\pi$ 
was shown to be\cite{{allen}}
\begin{eqnarray}
&&S(\omega,q\approx \pi,0)
\propto {m\over |\omega |}\delta
\left(\omega-\sqrt{v_s^2(q-\pi)^2+m^2}\right),\nn
&&S(\omega,q\approx 0, 0)\propto {m^2q^2\over
s^3\sqrt{s^2-4m^2}}\ , 
\end{eqnarray}
where $s^2=\omega^2-v_s^2q^2$. The explicit expressions for the
structure factor around $(q,q_\perp)=(0,\pi),\ (\pi,\pi)$ are
complicated, but reveal the presence of incoherent two and three
particle continua, respectively.
We will now show that the results obtained in the large-$N$ limit
are qualitatively incorrect. The reason for this failure of the
large-$N$ approach is that 
it entirely neglects the existence of topological kinks interpolating
between the two degenerate ordered ground states. 
Extrapolation of the large-$N$ results to lower values of $N$
shoud be done with caution because
the spectrum of the $O(N)$ Gross-Neveu 
model is very sensitive to the value of $N$ \cite{karow}.

\subsection{Exact results}
We now determine the dynamical structure factor using exact results on
formfactors in the Sine-Gordon model \cite{karowski,smirnov,lukyanov}.
We start with the case $q_\perp=0$, $q\approx 0$. The smooth component
of the sum of the two spin densities is expressed in terms of the
Sine-Gordon models as follows 
\begin{eqnarray}
S^z_1(x)+S^z_2(x)\bigg|_{\rm smooth}&\propto&
\partial_x\varphi_{+}+\partial_x\varphi_{-}\ .
\end{eqnarray}
This is nothing but the sum of the temporal components of the
current operators in the two Sine-Gordon models ($j^0_++j^0_-$). We
are interested in the structure factor, i.e., 
\bea
S(\omega,q\approx 0,0)
&\propto&{\rm Im} \sum_{\sigma=\pm}i \int_{-\infty}^\infty dx \int_0^\infty
dt~ e^{i(\omega+i\varepsilon)t-ivqx}\nn
&&\times\left\langle[j_\sigma^0(x,t)~,~
j_\sigma^0(0,0)]\right\rangle\ ,
\label{correl}
\eea
where $v_s$ is the velocity of the excitations. We express
\r{correl} in the spectral representation using our knowledge of a
complete set of states in terms of (anti) soliton scattering states. 
Energy and momentum are parametrised in terms of the rapidity variable
$\theta$ as
\begin{equation}
p=m\sinh\theta \qquad \epsilon=m\cosh\theta\ ,
\end{equation}
where $m$ is the mass of the four elementary excitations. 
The resolution of the identity is given
by
\begin{equation}
1\!=\!\sum_{n=0}^\infty\sum_{\alpha_i}\!\int
{d\theta_1\cdots d\theta_n\over (2\pi)^n n!}
\ket{\theta_n\cdots\theta_1}_{\alpha_n\cdots\alpha_1}
{}^{\alpha_1\cdots\alpha_n}\bra{\theta_1\cdots\theta_n},
\label{id}
\end{equation}
where $n$ is the number of particles and $\alpha_i\in\{s_\pm,\bar{s}_\pm\}$
specifies their respective ``flavour'' (soliton or antisoliton in $+$ or
$-$ sector). Inserting \r{id} in \r{correl} and using Poincar\'e
invariance yields
\begin{eqnarray}
&&S(\omega,q\approx 0,0) \propto\nn
&&-2\pi\ {\rm Im}\sum_{n=0}^\infty\sum_{\alpha_i}\int {d\theta_1\cdots
d\theta_n\over (2\pi)^n n!} |
F_{j^0}(\theta_1\cdots\theta_n)_{\alpha_1\cdots\alpha_n} |^2\cr 
&&\times\left[
{\delta\left( v_sq-m\sum_j\sinh\theta_j\right)
\over\omega-m\sum_j\cosh\theta_j+i\varepsilon}
-{\delta\left( v_sq+m\sum_j\sinh\theta_j\right)
\over\omega+m\sum_j\cosh\theta_j+i\varepsilon}
\right]\!,\nn
\label{correlff}
\end{eqnarray}
where $F_{j^0}(\theta_1\cdots\theta_n)_{\alpha_1\cdots\alpha_n}$
is the Sine-Gordon current form factor
\begin{equation}
F_{j^0}(\theta_1\cdots\theta_n)_{\alpha_1\cdots\alpha_n}\equiv
\bra{0}j^0(0,0)\ket{\theta_n\cdots\theta_1}_{\alpha_n\cdots\alpha_1}\
. 
\end{equation}
We note that an $n$-particles state only contributes to \r{correlff}
above the $n$-particle threshold, i.e. $s^2=\omega^2-v_s^2q^2\geq
n^2m^2$. Thus, at low energies $s^2\leq 16m^2$ only two-particle
states contribute. The corresponding formfactor is \cite{karowski}
\bea
&&F_{j^0}(\theta_1,\theta_2)_{s\sb}=-2m\sinh\left(
{\theta_1+\theta_2\over 2}\right)f(\theta_1-\theta_2)\ ,
\label{ffj0}\nn
&&f(\theta)=i{\sinh\theta/2\over 2\pi}\nn
&&\times
\exp\left(\int_0^\infty\!\! d\kappa{\sin^2\left({\kappa\over
2}(\theta-\pi i)\right) 
\over\kappa~{\rm sh}(\pi\kappa)}\left[{\rm th}\left({\pi\kappa\over 2}
\right)-1\right]\right).
\label{fmin}
\eea
After performing the $\theta$-integrations we obtain
\bea
&&S(\omega,q\approx 0,0) \propto {m^2v_s^2q^2 ~|f(2\theta(s))|^2
\over s^3\sqrt{s^2-4m^2}}\ ,
\label{strucfac0}
\eea
where $\theta(s)={\rm arccosh}\left({s\over 2m}\right)$ and
$4m^2<s^2<16m^2$. As we already mentioned, the result \r{strucfac0} is
exact as long as $s^2<16 m^2$. For larger energy transfers there are
(small) corrections due to four, six, eight etc particle states. These
can be calculated in the same way as the two-particle contribution.
Approaching the threshold $s=2m$ from above, 
\r{strucfac0} goes to zero like $\sqrt{s-2m}$.

The result (\ref{strucfac0}) has the same structure as the one
obtained in the large-$N$ approximation. We note that the vanishing of
the structure factor for $q=0$ ($S(\omega,0,0)=0$) reflects the fact
that the z-component of spin is a conserved quantity.

Next, we consider the structure factor at $(q\approx 0,\pi)$. The
smooth component of the difference of spin densities is
\bea
S^z_1(x)-S^z_2(x)\bigg|_{\rm smooth}
&\propto& \partial_\tau\varphi_{+}-\partial_\tau\varphi_{-}\ .
\end{eqnarray}
This is precisely the difference of the spatial components of 
the currents in the two Sine-Gordon models ($j^1_+-j^1_-$). 
Using the exact two-particle formfactor we obtain the leading
contribution to the structure factor
\be
S(\omega,q\approx 0,\pi) \propto {m^2\omega^2 ~|f(2\theta(s))|^2
\over s^3\sqrt{s^2-4m^2}}\ ,
\label{strucfacpi}
\end{equation}
where $f(\theta)$ is given by \r{fmin} and again
$4m^2<s^2<16m^2$. Note that the structure factor does not 
vanish for $q\to 0$ as the magnetisation difference between chains is
not conserved. This is due to the fact that our starting point does {\sl
not} have O(4) symmetry: after the duality transformation we obtain
an O(4) symmetric Lagrangian, but correlation functions transform
nontrivially. This result is of course expected, since the interchain
interaction must break the $O(4)\sim SU(2)\times SU(2)$ down to
$SU(2)$.

Finally, we examine the structure factor at $(q\approx \pi,0)$ and
$(q\approx \pi,\pi)$. The bosonised forms for the staggered components
of the sum and difference of the spin densities are found to be

\begin{eqnarray}
S^z_1(x)+S^z_2(x)\bigg|_{\rm stagg}
&\propto& \cos\sqrt{\pi}\Phi
\cos\sqrt{\pi}\Theta,\nn
S^z_1(x)-S^z_2(x)\bigg|_{\rm stagg}
&\propto& \sin\sqrt{\pi}\Phi
\sin\sqrt{\pi}\Theta\ ,
\label{spinpi2}
\end{eqnarray}
where $\Phi=(\varphi_++\varphi_-)/\sqrt{2}$ and
$\Theta=(\theta_+-\theta_-)/\sqrt{2}$.
At present it is not known how to calculate formfactors for the
operators appearing in \r{spinpi2} as they involve both the field and
the dual field. However, it is still possible to determine the
qualitative behaviour of the structure factor. From \r{spinpi2} it is clear
that the structure factor involves the calculation of formfactors of
operators
\bea
&&[{\rm cos\ or\ sin}]\left(\sqrt{\pi\over 2}\varphi_+\right)
[{\rm cos\ or\ sin}]\left(\sqrt{\pi\over 2}\theta_+\right)\cr
\times&&
[{\rm cos\ or\ sin}]\left(\sqrt{\pi\over 2}\varphi_-\right)
[{\rm cos\ or\ sin}]\left(\sqrt{\pi\over 2}\theta_-\right)\ .
\label{op}
\eea
These formfactors are obviously products of formfactors in
the two Sine-Gordon models. Let us therefore concentrate on the $+$
sector for the time being. It was shown in \cite{delfino} that the
operators $\cos\sqrt{\frac{\pi}{2}}\theta_+$ and
$\sin\sqrt{\frac{\pi}{2}}\theta_+$ in the Sine-Gordon model with
coupling constant $\beta=\sqrt{8\pi}$ have fermionic character and
thus have nontrivial formfactors with one-soliton states. On the other
hand, we know from \cite{karowski} that
$\cos\sqrt{\frac{\pi}{2}}\varphi_+$ and
$\sin\sqrt{\frac{\pi}{2}}\varphi_+$ 
are of bosonic character. We therefore conclude that $+$ part of the
operator \r{op} has fermionic character. This implies that it couples 
only to states with at least one (anti) soliton. An analogous statement
holds true for the $-$ sector, so that the leading contribution to the
structure factor comes from two-particle states. In other words no 
coherent one-particle excitation exists.

{From} the above results for the dynamical struture factor we deduce
that the low-lying excitations are described in terms of a gapped
two-particle scattering continuum. As we have mentioned above, the
elementary excitations carry spin-$1/2$. This leads us to identify
them as {\sl massive spinons}.

\section{Summary}
\label{section:summary}

We have studied the effects of pure current-current interactions in a
frustrated two-leg spin ladder. We have shown that {\sl spinons},
which are gapless topological excitations propagating along decoupled
Heisenberg chains, survive as elementary excitations in the frustrated
ladder, but acquire a finite mass gap. We have given an interpretation
of these massive spinons as quantum dimerisation kinks. The kinks are
deconfined and, in all physical states, appear only in pairs. 
As a result their contribution to the dynamical structure factor is
entirely incoherent. Our findings bear a strong resemblance to those
of \cite{NT}. 

We believe that our results not only apply to the ladder \r{ham},
but with some modifications also to the zigzag ladder \r{hamilspin}.
As discussed above, in the zigzag case there is a twist term in
addition to the current-current interaction. We conjecture, that the
effect of the twist term is merely to shift the minimum of the
two-spinon continua at $(q=\pi,0)$ and $(q=\pi,\pi)$ to
incommensurate wave numbers, i.e. to $(q=\pi+\delta,0)$ and
$(q=\pi+\delta,\pi)$, where $|\delta|\ll 1$. Such a picture is
consistent with what is known from numerical studies
\cite{wa,bursill,armando,andreas} and also fits well to what one would expect
on the basis of an (uncontrolled) extrapolation of the results for
$\delta={\cal O}(1)$ \cite{sush,brehmer} to $|\delta|\ll 1$.

Coming back to the twistless chain \r{ham}, it should be pointed out
that its ground state and excitations have been previously studied for 
the special case $J_\times=J$
\cite{bg,xian} (``Bose-Gayen model''). In this case, the
Hamiltonian \r{ham} exhibits an enlarged (local) symmetry, related
to the interchange ${\bf S}_1(n) \leftrightarrow {\bf S}_2(n) $
at {\sl arbitrary} rung $n$, and decouples into two commuting parts
describing either an array of entirely decoupled on-rung singlets
or an effective S=1 chain \cite{xian}. In both cases, the ground state
belongs to the universality class of the 
(undimerised) Haldane spin liquids with
the spin-1 massive magnons being coherent elementary 
excitations \cite{SNT,NT}. 
This is in marked contrast 
with our finidings for
$J_\times = \frac{1}{2} J_{\perp}\ll J$ and implies the 
existence of a crossover between the
two regimes at some intermediate coupling. 

It should be understood that the region where the marginally perturbed
ladder ($\lambda_2 = 0$) and the Bose-Gayen model
start overlapping, i.e. the vicinity of the point
$J_{\perp} = 2J_\times = 2J$,
is not accessible within our continuum approach, based on the assumption
that $J_{\perp}, J_\times \ll J$. Staying on the line
$J_{\perp} = 2J_\times$ and increasing $J_\times$ would enforce the amplitude
of the current-current perturbation ($\lambda_1$) to increase, in which
case no reliable conclusions are available. 
On the other hand, one can start approaching the Bose-Gayen regime
by keeping $J_{\perp}$ fixed and increasing $J_\times$. In this case
one inevitably deviates from the line $J_\perp=2J_\times$, and
that gives rise to the appearance of the strongly relevant perturbation
$\lambda_2 {\bf n}_1\cdot {\bf n}_2$. The latter 
introduces an extra potential,
\bea
{\cal U} &\sim& \lambda_2 [2 \cos \sqrt{2\pi}
\left(\varphi_+ - \varphi_- \right) - 
 \cos \sqrt{2\pi}\left(\varphi_+ + \varphi_- \right) \nonumber\\
&&\qquad+ \cos \sqrt{2\pi}\left(\theta_+ - \theta_- \right)],
\eea
that couples the two Sine-Gordon models (\ref{SG}), removes the $Z_2$
degeneracy between the two dimerised ground states and thus leads
to soliton confinement. The soliton-antisoliton pairs
start forming triplet and singlet massive bound states and transform to
coherent single-particle excitations. 
If the deviation from the line $J_\perp=2J_\times$ is large enough,
the $\lambda_2$ perturbation takes over, and the effective low-energy
field theory becomes that of four Majorana fermions, with a mass
term
$$
\propto i\lambda_2 
\left(\sum_{a=1}^3\psi_a\psib_a-3\psi_0\psib_0\right)
$$
as it is the case for the standard (nonfrustrated) ladder \cite{SNT}.
It is therefore
tempting to speculate that
the two massive Haldane phases on the both sides of the line $J_{\perp}
= 2J_\times$ can be smoothly connected with those of the Bose-Gayen model.
This, however, does not exclude the existence of other phases in the
3-parameter space of the model \r{ham}.

As discussed in Refs. \cite{conf,brehmer},
a similar soliton confinement scenario is realized 
if one adds an explicit dimerisation to the zigzag
Hamiltonian (\ref{hamilspin}).

\begin{center}
{\bf Acknowledgements}
\end{center}
We thank Armando Aligia, Vladimir Rittenberg and Alexei Tsvelik for
useful discussions. F.H.L.E. and A.A.N. are grateful to the
Physikalisches Institut der Universit\"at Bonn, where part of this
work was carried out, for financial support and hospitality.
F.H.L.E. is supported by the EPSRC under grant AF/98/1081.
D.A. would like to thank le Fonds FCAR du Qu\'ebec for
financial support.

\end{document}